\documentclass[3p,times]{elsarticle}

\usepackage{ecrc}

\volume{00}

\firstpage{1}

\journalname{Nuclear Physics A}

\runauth{}

\jid{nupha}

\usepackage{amssymb}
\usepackage[figuresright]{rotating}

\begin{document}

\begin{frontmatter}

%% Title, authors and addresses

%% use the tnoteref command within \title for footnotes;
%% use the tnotetext command for the associated footnote;
%% use the fnref command within \author or \address for footnotes;
%% use the fntext command for the associated footnote;
%% use the corref command within \author for corresponding author footnotes;
%% use the cortext command for the associated footnote;
%% use the ead command for the email address,
%% and the form \ead[url] for the home page:
%%
%% \title{Title\tnoteref{label1}}
%% \tnotetext[label1]{}
%% \author{Name\corref{cor1}\fnref{label2}}
%% \ead{email address}
%% \ead[url]{home page}
%% \fntext[label2]{}
%% \cortext[cor1]{}
%% \address{Address\fnref{label3}}
%% \fntext[label3]{}

\dochead{}
%% Use \dochead if there is an article header, e.g. \dochead{Short communication}

\title{Hard Probes 2010: Experimental Summary}

%% use optional labels to link authors explicitly to addresses:
%% \author[label1,label2]{<author name>}
%% \address[label1]{<address>}
%% \address[label2]{<address>}

\author{Federico Antinori}

\address{Istituto Nazionale di Fisica Nucleare, Sezione di Padova, via Marzolo 8, I-35131 Padova (Italy)}

\begin{abstract}
A (personal) experimental summary of the Hard Probes 2010 conference is presented.
%% Text of abstract
\end{abstract}

\begin{keyword}
hard probes \sep nucleus-nucleus collisions \sep RHIC \sep LHC \sep jet quenching \sep heavy flavours \sep parton energy loss
%% keywords here, in the form: keyword \sep keyword

%% MSC codes here, in the form: \MSC code \sep code
%% or \MSC[2008] code \sep code (2000 is the default)

\end{keyword}

\end{frontmatter}

%%
%% Start line numbering here if you want
%%
%%\linenumbers

%% main text

\section{Introduction}
This has been a very special edition of the Hard Probes conference: the first one after the start of LHC operation and the last one before the first Pb-Pb data from the LHC. It is not surprising, therefore, that the LHC took a prominent position in the programme. At the same time, new results keep coming out of RHIC, a clear indication of the continued liveliness of the ultrarelativistic heavy-ion programme at BNL a decade after the machine turn-on.
The present write-up is not meant to provide a comprehensive summary of the rich collection of high quality experimental contributions presented at the conference -- not enough space and in any case no way to do justice to the original detailed content, for which the individual contributions are essential reading -- but should rather be regarded as a sort of travel guide to a selection of sights from the conference material, biassed -- of course -- through personal taste and prejudice.
Even the material after selection would have been just too much to include the figures illustrating the results. Since, on the other hand, the present paper is not meant to live an independent life, separated from the rest of the proceedings, the inclusion of figures would have represented a useless duplication anyway. Therefore, I have decided not to include any, and the reader is encouraged to refer to the original contributions, where the results are properly qualified.
For a bird's eye view of the main plots relative to the material included in the present digest, the reader is referred to the slides of my oral contribution  \cite{slides}, which were mostly assembled from the original material presented at the conference.

\section{The LHC is here}
The first highlight of this conference is the LHC. The accelerator is working beautifully, the experiments are performing to specifications and producing good physics.
We saw many examples at this conference of both the level of understanding of the experimental apparatus (for example \cite{koch, dolejsi}), already remarkable at such an early stage, and the level of agreement across the experiments for the first pp measurements, almost perfect \cite{safarik}.
Proton-proton results relevant to heavy-ion physics are among the main highlights of this conference.

The first Pb-Pb collisions are just around the corner, at the reduced energy of 2.76 TeV (corresponding to Pb ions accelerated at the same magnetic rigidity of protons of 3.5 TeV, as currently used in the LHC).
The expected luminosity for the first run is of the order of $10^{25}\rm{cm}^{-2}\rm s^{-1}$, about two orders of magnitude below nominal (compared to the nominal Pb running scheme, in the early running scheme for 2010 roughly one order of magnitude reduction in luminosity is due to the reduced number of bunches  and roughly one order of magnitude is due to reduced focussing). The expected rate of minimum bias nuclear interactions is 50 - 100 Hz, leading to the use of very open triggers. This first run is expected to provide an integrated luminosity of a few $\mu \rm b^{-1}$ per experiment. Such statistics, although representing only a small fraction of the collaborations' target (several $\rm{nb}^{-1}$), should already allow to perform key measurements of the global properties of the system (multiplicity, elliptic flow, Hanbury Brown -- Twiss interferometry, bulk strangeness production), to access nuclear modification factors, particle correlations and  identified particle spectra (with a transverse momentum reach dependent on statistics) and to get a first glimpse of jet, quarkonia and heavy flavour physics.

\section{Quarkonia}
Progress is still being made towards a better
theoretical understanding of $J/\psi$ production in proton-proton collisions.
The production of the $J/\psi$ at Tevatron has been puzzling the community for more than a decade \cite{CDF,yellow}.
Calculations based on the use of leading-order pQCD diagrams for the production of $c\overline{c}$ pairs in colour singlet seriously underestimate the $J/\psi$ cross-section at Tevatron. The introduction of the contribution from colour octet channels allows to bring up the theoretical cross section, back in line with experiment, but at the expense of the prediction of a large polarisation at high transverse momentum, not observed in the data.
A recent theoretical study, discussed at this conference \cite{maltoni}, indicates that once the dominant next-to-next-to-leading order (NNLO) diagrams are included in the calculations, the bulk of $J/\psi$ production can actually be accounted without invoking colour-octet mechanisms, which could help reconcile theory with the polarisation measurements. Will the introduction of the higher order terms finally lead to a consistent description of quarkonium production at colliders? On the other hand, if NNLO contributions cannot be neglected, what about higher orders? This continues to be a fascinating subject, with further progress to be expected in the future.

Meanwhile, high quality data are pouring out of the LHC: a beautiful dimuon landscape was shown by CMS, who presented the mass spectrum ranging from the $\rho/\omega$ peak all the way up to the Z$^0$ \cite{castello, woehri}. Preliminary $p_T$-differential $J/\psi$ cross sections were presented by ALICE \cite{arnaldi}, ATLAS \cite{maiani} and CMS \cite{woehri}, with very good agreement across the experiments. The prompt and non-prompt contributions to the $J/\psi$ samples are already being disentangled by separating the dimuon vertex from the primary vertex: CMS has extracted the fraction of $J/\psi$ from B decays as a function of transverse momentum, finding values compatible with the ones from Tevatron \cite{lourenco}. CMS also presented a $p_T$-differential $\Upsilon$ cross-section \cite{lourenco}: quarkonium physics at the LHC is quickly reaching full speed.

Turning to nuclear collisions: new quarkonium results were presented at this conference by the RHIC experiments. The STAR collaboration has extracted a $\Upsilon$ signal in Au-Au, based on 0.3 nb$^{-1}$ from the 2007 run, and is even quoting a preliminary value for the nuclear modification factor $R_{AA}$, although with large uncertainties \cite{reed, xie}. A further 1.4~nb$^{-1}$ is already on tape \cite{reed}, and more news on $\Upsilon$ production at RHIC are to be expected in the near future.  The PHENIX collaboration has studied the $J/\psi$ central-to-peripheral nuclear modification factor $R_{CP}$ in dAu collisions as a function of rapidity, observing a decrease, when going from backward to forward rapidities \cite{wysocki}. The centrality of the dAu collision is established based on the amplitude measured in the experiment's Beam-Beam Counters. Compared to $R_{dAu}$, the dAu $R_{CP}$ has a significantly reduced systematic uncertainty. The result is compatible with shadowing/saturation expectations. The interpretation, however, is complicated by the fact that the nuclear modification factor $R_{dAu}$ for "peripheral" dAu collisions is itself significantly different from unity, so that these collisions cannot be directly used as an approximation for pp.
A quantum jump is expected with the turn-on of Pb-Pb collisions at the LHC later this year: with O(100) $c\overline{c}$ pairs expected per central Pb-Pb event, charmonium physics will enter a new era, in which the interplay of initial suppression and recombination could open the way to a novel phenomenology.

\section{Two-particle correlations}
Triangular flow came under the spotlight in 2010, and the subject was reviewed in detail at this conference \cite{roland}: event-by-event "triangularity" fluctuations in the participants' distribution could lead to a triangular modulation (v$_3$) in the azimuthal distributions. This "participant triangularity" may actually explain the two main structures observed in two-particle correlations at RHIC: the near-side ridge and the double peak on the away-side. Could this be the end of the Mach cone interpretation of the away-side structure? Or would this be at odds with the evidence from STAR that the away-side structure is conical \cite{conical}, and therefore cannot be accounted for solely by initial state fluctuations?
Better understanding is needed to clarify the relative importance of geometrical fluctuations in the initial state geometry and medium response effects in shaping the two-particle correlations landscape.
Rather puzzling, in this context, is the observation of a ridge-like correlation on the near-side in high multiplicity pp collisions reported by CMS earlier this year \cite{ppridge}. What is the relationship, if any, with the nucleus-nucleus ridge?

An interesting perspective in the particle correlations sector is the possibility of studying correlations involving heavy flavour signals (non-photonic electrons, reconstructed D, J/$\psi$,...), which would bring information on the B cross-section / contamination and may even provide insights on the medium response to heavy partons. Certainly another area to be closely watched in the near future.

\section{Jets}
With Pb-Pb collisions at the LHC just around the corner, jets have had a particularly prominent place in this edition of Hard Probes. The RHIC collaborations have presented their results, putting forward rather contrasting pictures. The PHENIX collaboration, extracting jets using gaussian filtering, presented results from the analysis of Cu-Cu collisions, with jet $R_{AA}$ values in line with those found for single particles ($\pi^0$) \cite{lai}. On the other hand, the STAR collaboration, reconstructing jets with the $K_T$ and anti-$K_T$ algorithms, observes, in Au-Au collisions, significantly less suppression than for the case of inclusive particle production \cite{putschke}.
STAR has also presented an interesting background correction study, that utilizes simulated jets embedded into real Au-Au events in order to extract a distribution of the energy fluctuations to be used in the unfolding corrections \cite{jacobs}.
PHENIX has studied the centrality dependence of jet-jet azimuthal correlations as a function of centrality in Cu-Cu, observing no evidence of broadening \cite{lai}, while STAR, studying the distribution of the energy within the jet as a function of the radius in Au-Au, finds evidence for substantial broadening \cite{putschke}.
Both collaborations have reported at this conference evidence for softening of the fragmentation function on the away-side: STAR based on the study of jet-hadron correlations \cite{bruna} and PHENIX measuring the fragmentation function of jets recoiling against photons \cite{connors}.
While our understanding of jet production in nucleus-nucleus collisions from RHIC may still be somewhat unsettled, jet physics is expected to be cleaner at the LHC, with much larger jet cross-sections and better jet capabilities in the experiments. Things are already moving fast in pp: the understanding of jet energy scale calibrations is precociously good (e.g.: \cite{andreazza}), double-differential cross-sections out to several hundreds GeV are already available (e.g.: \cite{fullana}), and we even saw results on the cross-sections for b-tagged jets \cite{chiochia}.
Jets in the 100 GeV range, which should be comfortably accessible at the LHC, would stand out from the background distinctly also in Pb-Pb events and should allow for clear jet quenching measurements.
On the other hand, in order to understand the fate of the lost energy, we will have to push down in reconstructed particle momenta as much as possible, into the backgrounds again, where life at the LHC is not going to be simple.
We will have to disentangle the interplay of a fluctuating background with as yet unknown modifications of the jet shapes. Background fluctuations can modify the way a given jet is reconstructed. The effect of such back-reactions was discussed at this conference, within a detailed study of the expected performance of some of the main jet reconstruction algorithms for central Pb-Pb collisions at the LHC \cite{soyez}. According to this study, the anti-$K_T$ algorithm and an improved version of the Cambridge/Aachen algorithm (modified with the addition of filtering) show the least sensitivity to back-reaction (although for the latter, this seems to be due to a fortuitous cancelation between a negative back-reaction and a positive bias from the filtering \cite{soyez}).
While progress is clearly being made, the issue of the best approach to jet reconstruction in heavy-ion collisions remains far from settled. Contrasting views were voiced at this conference, with signs of incipient polarisation in this area. Such a plurality of views has also positive aspects as we enter the unchartered territory of the LHC: it will certainly be very interesting to look at jets using different approaches and compare the results, and it may turn out that in the end a combination of different tools is what we need, with -- for instance -- the more rigid, cone-like approaches better suited for quenching measurements and more "organic" algorithms better adapted to studies of jet softening/broadening.
And of course the flavour capabilities of the LHC experiments will come into the game, allowing heavy flavour tagging of jets and identified particle fragmentation studies. The future of the jet sector of our field promises to be very rich.

\section{Energy loss}
Significant progress in the comparison of the different energy loss formalisms has been made in the framework of the TECHQM collaboration \cite{techqm}: in order to get rid of the differences introduced by variations in the modeling of the medium, energy loss calculations based on four different formalisms have been performed for a standardized, static medium of fixed length and density ("the brick"), and the outgoing parton spectra have been compared in two different conditions: same density and same suppression \cite{marco}. Even for this simplified case, significant differences are observed between the calculations, indicating that the effects of the particular approximations used in the different formalisms are sizeable.
More work is needed to put energy loss calculations on a firm quantitative basis for comparison with data, and, with the LHC approaching, this issue is becoming more and more pressing.

On the experimental side, an interesting study was presented by PHENIX \cite{gong}. They showed that perturbative Quantum Chromo-Dynamics (pQCD) based energy loss models have difficulty in reproducing simultaneously the PHENIX results on $\pi^0$ $R_{AA}$ and $v_2$ and studied the behaviour of the suppression as a function of the $\pi^0$ emission angle relative to the reaction plane. For transverse momenta above 5 - 6 GeV/$c$, they observe an interesting scaling pattern:
when the measured values of $R_{AA}$ are plotted as a function of a value of the path-length integral calculated assuming Anti de Sitter/Conformal Field Theory (AdS/CFT)-like jet-medium interactions (i.e. $\propto L^3$ instead than $\propto L^2$ as for pQCD) and Colour Glass Condensate (CGC) initial geometry fluctuations, the points fall on the same curve for all emission angles. Certainly an intriguing regularity.

\section{Heavy flavour}
With in-medium energy loss expected to be dependent both on the mass and on the colour charge of the propagating parton, the study of the nuclear modification of heavy flavour production promises to provide important insights on the properties of the medium. This area has attracted a lot of attention in recent years, but has been plagued by a long-standing discrepancy between STAR and PHENIX \cite{STARNPE,PHENIXNPE} (and between STAR and fixed-order-next-to-leading-log -- FONLL \cite{FONLL} -- pQCD predictions \cite{STARNPE}) on charm production in Au-Au collisions.
Starting from the 2008 run the STAR collaboration has removed the Silicon Vertex Tracker and the Silicon Strip Detector, significantly reducing the background to the non-photonic electron (NPE) signal. The results for the STAR NPE transverse momentum spectra in Au-Au collisions \cite{xie} are now compatible with those of PHENIX. The analysis of data collected by STAR in 2005 with the silicon detectors still in place also yields compatible results. The original 2003 data on which the STAR publication \cite{STARNPE} was based are now being re-analysed. The STAR value for the cross section, which is dominated by the reconstructed D meson results, is unchanged, so the discrepancy with PHENIX for the total $c\overline{c}$ cross section is still unsolved. The discrepancy between the two experiments on the $p_T$-differential cross sections in pp \cite{PHENIXpp,STARNPE} also still remains.

Turning to the nuclear modification factor,
the tension between the RHIC values for $R_{AA}$ and the pQCD-based quenching predictions for non-photonic electrons is still extant. Here, again, AdS/CFT predictions seem to do better than the pQCD-based ones \cite{horowitz}. More points for string theory? There is an astounding prediction from AdS/CFT in the heavy flavour sector at the LHC: the ratio
$R^c_{AA}(p_T)/R^b_{AA}(p_T)$ between the charm and beauty nuclear modification factors is predicted to level off at high transverse momenta to a value around 0.2 \cite{horowitz}, in contrast with the values around unity predicted by pQCD-based calculations (and generally expected based on the assumption that the effects of the b/c mass difference should disappear in the ultra-relativistic limit).
This prediction will soon be tested: charm/beauty separation in nucleus-nucleus collisions will finally be possible at the LHC, where the experiments are equipped with high resolution vertex detectors (heavy flavour vertexing capabilities are also expected to become available at RHIC in the future, with the planned addition of silicon pixel detectors to both the STAR and PHENIX layouts).
The potential fly in the LHC ointment, until the pA run, is represented by parton shadowing/saturation: for D mesons, for instance, care will have to be exercised at transverse momenta below 10 GeV, where initial state effects are expected to be important.
Waiting for Pb-Pb collisions, beautiful heavy flavour signals from the LHC are being extracted in pp: at this conference we were treated to heavy flavour electron \cite{masciocchi} and muon \cite{stocco} transverse momentum spectra, $D^0$ and $D^+$ transverse momentum spectra \cite{dainese}, and we saw the first signals of $D^*$ \cite{dainese,woehri}, $D_S$ \cite{dainese} and $B^\pm$ \cite{woehri}. The LHC experiments are performing to expectations, and the heavy flavour future is looking very bright.

\section{Conclusions}
We are at a very exciting crossroads in the field of ultrarelativistic nuclear collisions, with little over one month to go before the first Pb-Pb collisions at the LHC.
With a jump in collision energy of one order of magnitude, and later more, the LHC will usher in a host of new, abundant hard probes. The combination of cleaner signals and more reliable theoretical predictions will undoubtedly provide us with a powerful toolset to advance the quantitative understanding of the nucleus-nucleus initial state wavefunction, the in-medium energy loss and the medium response.

\section*{Acknowledgements}
The author would like to thank the organisers, for the honour and pleasure of summarising this important edition of the Hard Probes conference; Brian Cole,
Megan Connors, Andrea Dainese, David d'Enterria, Barbara Jacak, Carlos Louren\c{c}o, Jos\'e Guilherme Milhano, Gunther Roland, Karel \v{S}afa\v{r}\'ik, Carlos Salgado, Itzhak Tserruya, Thomas Ullrich, Urs Wiedemann and Wei Xie, for discussions, help and support; and the organisational staff at the conference, whose kind and friendly assistance was instrumental in making it possible for the author to carry through the daunting task of preparing the experimental summary.

%% The Appendices part is started with the command \appendix;
%% appendix sections are then done as normal sections
%% \appendix

%% \section{}
%% \label{}

%% References
%%
%% Following citation commands can be used in the body text:
%% Usage of \cite is as follows:
%%   \cite{key}         ==>>  [#]
%%   \cite[chap. 2]{key} ==>> [#, chap. 2]
%%

%% References with BibTeX database:

\bibliographystyle{elsarticle-num}
\bibliography{<your-bib-database>}

\begin{thebibliography}{00}

%% \bibitem must have the following form:
%%   \bibitem{key}...
%%
\bibitem{slides}
\begin{verbatim}
http://www.weizmann.ac.il/MaKaC/materialDisplay.py?contribId=175&sessionId=47&materialId=slides&confId=2
\end{verbatim}
\bibitem{koch} K Koch et al. (ALICE), this conference.
\bibitem{dolejsi} J Dolej\v{s}\'i et al. (ATLAS), this conference
\bibitem{safarik} K \v{S}afa\v{r}\'ik et al. (ALICE), this conference.

\bibitem{CDF} F. Abe et al. (CDF), Phys. Rev. Lett. 79 (1997), 578
\bibitem{yellow} N Brambilla et al., arXiv:hep-ph/0412158
\bibitem{maltoni} F Maltoni, this conference.
\bibitem{castello} R Castello et al. (CMS), this conference.
\bibitem{woehri} H W\"ohri et al. (CMS), this conference.
\bibitem{arnaldi} R Arnaldi et al. (ALICE), this conference.
\bibitem{maiani} C Maiani et al. (ATLAS), this conference.
\bibitem{lourenco} C Louren\c{c}o et al. (CMS), this conference.
\bibitem{reed} R Reed et al. (STAR), this conference.
\bibitem{xie} W Xie et al. (STAR), this conference.
\bibitem{wysocki} M Wysocki et al. (PHENIX), this conference.
\bibitem{roland} G Roland, this conference.
\bibitem{conical} B I Abelev et al. (STAR), Phys. Rev. Lett. 102 (2009), 052302
\bibitem{ppridge} V Khachatryan et al. (CMS), JHEP 09 (2010), 091
\bibitem{lai} Y S Lai et al. (PHENIX), this conference.
\bibitem{putschke} J Putschke et al. (STAR), this conference.
\bibitem{jacobs} P Jacobs et al. (STAR), this conference.
\bibitem{bruna} E Bruna et al. (STAR), this conference.
\bibitem{connors} M Connors et al. (PHENIX), this conference.
\bibitem{andreazza} A Andreazza et al. (ATLAS), this conference.
\bibitem{fullana} E Fullana Torregrosa et al. (ATLAS), this conference.
\bibitem{chiochia} V Chiochia et al. (CMS), this conference.
\bibitem{soyez} G Soyez et al., this conference.
\bibitem{techqm}
\begin{verbatim}
https://wiki.bnl.gov/TECHQM/index.php/Main_Page
\end{verbatim}
\bibitem{marco} M van Leeuwen, this conference.
\bibitem{gong} X Y Gong et al. (PHENIX), this conference.
\bibitem{STARNPE} B I Abelev et al. (STAR), Phys. Rev. Lett. 98 (2007), 192301
\bibitem{PHENIXNPE} A Adare et al. (PHENIX), Phys. Rev. Lett. 98 (2007), 172301
\bibitem{FONLL} M Cacciari et al., Phys. Rev. Lett. 95 (2005), 122001
\bibitem{PHENIXpp} A Adare et al. (PHENIX), Phys. Rev. Lett. 97 (2006), 252002
\bibitem{horowitz} W Horowitz, this conference.
\bibitem{masciocchi} S Masciocchi et al. (ALICE), this conference.
\bibitem{stocco} D Stocco et al. (ALICE), this conference.
\bibitem{dainese} A Dainese et al. (ALICE), this conference.


\end{thebibliography}

%% Authors are advised to use a BibTeX database file for their reference list.
%% The provided style file elsarticle-num.bst formats references in the required Procedia style

%% For references without a BibTeX database:

\end{document}